Electrical tuning of the spin-orbit interaction in nanowire by transparent ZnO gate grown by atomic layer deposition

Keiko Takase, Kouta Tateno, and Satoshi Sasaki
NTT Basic Research Laboratories, NTT Corporation, 3-1 Morinosato-Wakamiya, Atsugi, Kanagawa, 243-0198, Japan

Corresponding author: Keiko Takase
Email: keiko.takase.wa@hco.ntt.co.jp

We develop an InAs nanowire gate-all-around field-effect transistor using a transparent conductive zinc oxide (ZnO) gate electrode, which is *in-situ* atomic layer deposited after growth of gate insulator of $Al_2O_3$. We perform magneto-transport measurements and find a crossover from weak localization to weak antilocalization effect with increasing gate voltage, which demonstrates that the Rashba spin-orbit coupling is tuned by the gate electrode. The efficiency of the gate tuning of the spin-orbit interaction is higher than those obtained for two-dimensional electron gas, and as high as that for a gate-all-around nanowire metal-oxide-semiconductor field-effect transistor that was previously reported. The spin-orbit interaction is discussed in line with not only conventionally used one-dimensional model but also recently proposed model that considers effects of microscopic band structures of materials.

[Text]

   In recent years, there has been growing interest in making devices with new functionality based on spin-orbit interaction [1], which is a relativistic effect that couples the electron spin to its momentum. The spin-orbit coupling arises in the presence of a lack of bulk or structural inversion symmetry. The latter case is known as the Rashba spin-orbit coupling, which can be controlled by external electric fields. The controllability has stimulated researchers to apply this effect for a spin field-effect transistor (FET) [2], Majorana-based devices [3,4], and a topological computer [5].  With the increasing attention, electrical control of the spin-orbit interaction with high gate-bias efficiency and obtaining a large spin-orbit interaction with a low gate bias become increasingly important.

Recently, Liang et al. developed an InAs nanowire device with a surrounding gate structure using ion electrolyte, which demonstrates a large tuning of the spin-orbit interaction [6]. As a different approach for a high tuning of the spin-orbit interaction without ion liquid, we have developed devices such as a gate-all-around (GAA) InAs nanowire metal-oxide-semiconductor FET [7] and InSb nanowire with a nearby back-gate structure [8]. They enable to control the Rashba spin-orbit coupling more efficiently than InAs nanowires with standard MOS gate structures [9–13], a quasi-one-dimensional InGaAs/InP wire [14] and the two-dimensional conventional Schottky FETs fabricated from III-V semiconductor quantum wells [15,16].

On the other hand, while electrical fast control and application to optoelectronics reaching to THz regime are advanced in III-V semiconductor quantum dots employing strong spin-orbit interaction [17,18], such a challenge for III-V semiconductor nanowire FETs with large spin-orbit coupling are still developing. Some reasons are as follows. Firstly, there have been only a few reports on devices that can largely tune the spin-orbit interaction by the gate with metal-oxide-semiconductor (MOS) structure, which can work in a standard FET speed and can be further devised to reduce interface states. Secondly, standard MOS devices use metal gate but it screens high-frequency range. Therefore, for a future application to high-frequency like other quantum devices [18], it is important to demonstrate a III-V semiconductor nanowire device with a gate that has a high transmittance for GHz and THz range, and first necessary to investigate low-frequency response of electrical-field control of the spin-orbit interaction as was widely studied [6-16].

Here, we examine low-frequency gate response of the spin-orbit interaction with conventionally used magneto-transport method for an InAs nanowire FET with a GAA structure using *n*-doped conductive ZnO. The device employs an *in-situ* fabricated gate stack that can possibly reduce interface states. ZnO is known to have high transmission for a wide range of light from visible to far-infrared (THz) regime, which indicates that this material choice has a potential application for optoelectronics [19,20]. Moreover, our fabrication methods employing an atomic-layer deposition (ALD) growth of ZnO includes additional benefits as compared to normal metal gate electrode formation as detailed later. Our measurements demonstrate that the Rashba spin-orbit interaction is controlled by the low gate voltage with a high gate efficiency. Our nanowire device using *in-situ* grown high-frequency transparent ZnO/Al$_2$O$_3$ gate stack can contribute to a future prototype of the spin FET that can be applied in optoelectronics.

We use InAs nanowire grown by MOVPE method with a catalyst of gold nano particle [21]. Our nanowire has a wurtzite structure grown along *c*-axis with a hexagonal cross-sectional shape. We *in-situ* grow $Al_2O_3$ (6 nm) and conductive ZnO (20 nm) around InAs nanowire by atomic-layer deposition technique in a different chamber from that used for nanowire growth. Since ZnO and $Al_2O_3$ serve as gate electrode and gate insulator, we can omit one fabrication process of spin-coating resist that is usually required to protect contact area before gate-metal deposition [22]. This can reduce possible contamination and interface states [23–25] between gate insulator and conductive electrodes. Moreover, ALD growth method enables us to realize the surrounding gate electrode with uniform thickness more easily than using tilted angle deposition of normal metal. Therefore, yields for making perfectly covering GAA devices are much improved compared to the cases using metal deposition.

Conductive ZnO is formed by ALD using diethylzinc and water as precursors. By adjusting deposition temperature, we grow *n*-doped film with typical resistivity of $5 \times 10^{-3}$ Ω·cm. We also note that our ALD-grown ZnO film is transparent for a wide range of frequency (~ 10 THz) and the transmittance is higher (~ 90 %) when the film has higher specific resistivity. For this device, we placed importance on the fact that conductive ZnO serves as a perfectly covering GAA electrical gate in order to be satisfactorily used in electrical transport. Therefore we used $ZnO/Al_2O_3$ gate-stack with lower resistivity as in pure metal, and thus the gate-stack used this time has slightly lower transmittance (~ 60 % up to 10 THz) among the $ZnO/Al_2O_3$ films that we have checked (not shown here). This is quite different from our previous GAA device using pure metal gate (Ti/Au), because our previous metallic-gate GAA device is considered to screen high-frequency such as GHz and THz by nearly 100 % and thus the previous device needs to be improved for a high-frequency operation. However, in this study, we use direct-current operation of the spin-orbit interaction, and thus the ZnO film properties of THz transmission difference such as lower/higher (higher/lower) transmittance (conductivity) does not affect the gate controllability of the spin-orbit interaction.

Figures 1(a) to (g) show scanning transmission electron microscopy (TEM) images of a typical InAs nanowire coated with $Al_2O_3/ZnO$ grown by ALD. Figure 1(a) shows a high-angle annular dark field image, which demonstrates that InAs nanowire was coated with $Al_2O_3/ZnO$ of uniform thickness. Figures 1(b) to (d) show energy-dispersive X ray spectroscopy (EDS) images to be used for elemental analysis. These figures indicate that our device has no notable impurities or unintended admixtures of elements. Figure1(g) corresponds to the superposition of Figs. 1(c), (d) and (e). This clearly indicates that $ZnO/Al_2O_3$ layers are well grown around InAs nanowire by ALD technique. The overall schematic illustration of our device is shown in Fig. 1(h). The $ZnO/Al_2O_3$ coated nanowire is deposited on a pre-patterned metal strip. The ALD grown ZnO and Al$_2$O$_3$ films coat the whole InAs nanowire, but in the vicinity of the contact area, ZnO was wet-etched and then Al$_2$O$_3$ and surface of InAs nanowire were etched by Ar plasma to connect the source and drain metal electrodes. This procedure prevents from a parasitic channel that can be formed in the conductive ZnO film. We also checked the absence of the gate leak current within the gate bias window shown in the paper, which demonstrates that there is no parasitic channel. The transport measurements are done at temperature of 1.6 K.

Figure 2 shows gate-voltage ($V_g$) dependence of source-drain current ($I_{sd}$), which is measured for various source-drain voltage ($V_{sd}$) at 1.6 K. With increasing $V_g$, $I_{sd}$ drastically changes for all $V_{sd}$. Our FET shows that electrical current starts to flow at $V_g > 0$, indicating an enhancement-type behavior. We consider that the area in the vicinity of the contacts is not depleted at any bias and the Fermi level position can be different between that area and the gated region, since the Fermi level in InAs nanowire is pinned at the surface and thus it is very sensitive to the surface condition. The field-effect mobility for this device is estimated to be 120 cm$^2$/Vs. Subthreshold swing is about 30 meV/dec, which is slightly larger or comparable to those of nanowire FETs previously reported [7,8,26].

We next measure magnetoconductance to investigate whether the spin-orbit interaction is tuned by the ALD-grown ZnO gate electrode. Figure 3(a) shows magnetoconductance ($\Delta G$) as a function of magnetic field ($B$). Here we define $\Delta G$ as an increment from zero-field conductance. Like previous reports dealing with nanowire magneto-transport, we smoothen $G$ with respect to $B$ to reduce impact of universal conductance fluctuation and average $G$ for positive and negative $B$, both of which work to obtain better accuracy of fitting that we explain below. Figure 3(a) demonstrates that $B$ dependence of $\Delta G$ shows a dip to a peak structure with increasing $V_g$. This behavior is a crossover from weak localization to weak antilocalization, the latter of which is known to arise in the presence of strong spin-orbit interaction [27].

To estimate the spin-orbit interaction, we here use the one-dimensional model that has been widely used for diffusive transport in nanowires. We analyze our data coherently from weak localization to weak antilocalization regime with this model as has been done in many groups, while we mainly focus on the weak antilocalization regime where we can find that the spin-orbit interaction relevantly plays an important role. We discuss the relation of this model with other recently proposed models later. In one-dimensional disordered model, the fitting formula is given by [9],

$$\Delta G = -\frac{2e^2}{hL_g}\left[\frac{3}{2}\left(\frac{1}{l_\phi^2}+\frac{4}{3l_{so}^2}+\frac{W^2}{3l_B^4}\right)^{-1/2} - \frac{1}{2}\left(\frac{1}{l_\phi^2}+\frac{W^2}{3l_B^4}\right)^{-1/2}\right], \qquad (1)$$

where $e$ is electron charge, $h$ is the Planck constant, $l_\phi$ is the phase relaxation length, $l_{so}$ is the spin-orbit length, and $l_B$ is magnetic length given by $l_B = \sqrt{\hbar/eB}$ ($\hbar$ is the reduced Planck constant, i.e. $h$ divided by $2\pi$). $W$ is nanowire diameter and $W = 65$ nm. $L_g$ is the gate length and $L_g = 2.0$ μm. To satisfy a fitting range of small field, we fit our data within $-0.15$ T $< B < 0.15$ T, corresponding to $W \leq l_B$.

We note that our nanowire has a mean free path (~ 4 nm) smaller than $W$ and that $W < l_\phi < L_g$ is satisfied. Thus requirements of using the one-dimensional diffusive model for weak antilocalization is reasonable. As is shown in Fig. 3(a), our data are nicely fitted with a parameter set of only $l_{so}$ and $l_\phi$ using Eq. (1). Figure 3(b) shows $l_{so}$ and $l_\phi$ plotted as a function of $V_g$. We find that $l_{so}$ is significantly decreased with $V_g$, indicating that the spin-orbit interaction is tuned by the gate electric field.

Figure 4(a) compares $l_{so}$ for our FET with those for differently structured FETs using InAs nanowire. We note that all the devices use the same one-dimensional diffusive model to obtain $l_{so}$. The device types are categorized into two groups. One is a so-called GAA structure [6,7], in which gate insulator and gate electrodes coaxially surrounds a nanowire. The spin-orbit interaction examination for nanowire with GAA structure was pioneered by Liang et al. [6], who used electrolytes for gate insulator and achieved the very high gate efficiency of the spin-orbit interaction. This gate structure excels in the efficiency than ours using metal-oxide gate structure. However, the gate operation of electrolytes generally needs enough time and thermal cycles to stabilize the ion distribution in the insulator, and so our metal-oxide GAA structures using high-k gate insulator have advantage for higher frequency application. The other is a structure with standard gate such as back-, top- or side-gate [9–12], which applies electric fields in single or dual direction to the nanowire deposited on $SiO_2$/Si wafer.

As shown in Fig.4 (a), we can find that the data obtained for the GAA type FETs show a sharp $V_g$ dependence in a few volts range, while those for back-, top-, or side-gated FETs show a flat $V_g$ dependence. This clearly indicates that the spin-orbit interaction is significantly varied by the gate tuning for the GAA FETs. We also note that the gate bias windows of sharp $l_{so}$ vs. $V_g$ is more positive for ZnO-GAA device than that for metallic-GAA device that we previously developed, while the devices geometry and threshold voltages of the two devices are similar. We consider that this is associated with the microscopic carrier distribution difference in nanowires, which can be affected by impurities and stacking faults related to nanowire material growth as well as interface states between $Al_2O_3$ and InAs, the latter of which may be influenced by difference between metallic gate and conductive ZnO gate. In addition, we speculate that this $V_g$ shift for $l_{so}$ might be related to spin-relaxation caused by impurities based on Elliot-Yafet mechanism (spin flip that happens probabilistically due to impurity) [28] [29], which occurs in the presence of the spin-orbit interaction and is different from Rashba-type spin precession. Since the mobility in the present device is about ten times smaller than that in the previous device, the present device is affected by impurity scattering more strongly, which indicates that spin-relaxation due to impurities (Elliot-Yafet mechanism) can more strongly occur and can compete with the Rashba-type spin precession (D'yakonov-Perel mechanism [30]). This competition may bring the shift of the gate voltage where the Rashba spin-orbit interaction manifests itself as in the form of the weak antilocalization.

Then we estimate the strength of the spin-orbit interaction using $l_{so}$. The Rashba coupling parameter $\alpha_R$ is defined in the Hamiltonian, $\alpha_R \cdot (\boldsymbol{\sigma} \times \mathbf{k})$, where $\boldsymbol{\sigma}$ is the Pauli spin matrices and $\mathbf{k}$ is the electron wavevector. It is also known that $\alpha_R$ is given by a simple form of $\alpha_R = \hbar^2/2m^* l_{so}$, where $m^*$ is effective mass of 0.023 $m_e$ ($m_e$: electron mass). Using this, we obtain the Rashba coupling parameter for our ZnO-GAA device. Figure 4(b) shows $\alpha_R$ for the GAA-type devices. We find that ZnO- and our previous metal-GAA devices have a sharp slope with a similar magnitude in $\alpha_R$ as a function of $V_g$, indicating that gate efficiency for tuning the spin-orbit interaction is comparably large. We note that the gate efficiency is larger than InAs nanowires with standard MOS structures [9–12] and one-dimensional [14] and two-dimensional [15,16] devices fabricated from III-V semiconductor quantum wells.

In the above discussion, we have extracted the Rashba spin-orbit interaction using one-dimensional diffusive model, which was used in various experiments [6–13] and originally developed to be used regardless of material. This fact also indicates that electrical transport in nanowire FETs has been generally diffusive, with the exception of a quasi-ballistic channel in InSb nanowire [31], which was recently demonstrated and analyzed by the three-dimensional ballistic model that they established for their experiments. Therefore, theories dealing with InAs nanowire have focused on diffusive transport, and recently examined the spin-orbit interaction having assumed how electrons propagate in nanowire with specific crystal structures. Hereafter, we refer to the other models that are recently proposed.

Kammermeier et al. reported models that can analyze weak localization and weak antilocalization to extract the spin-orbit interaction for InAs nanowire with zinc-blend (111) and (100) [32,33] and wurtzite (0001) structures [34]. They assumed various shapes in which electron can propagate in nanowires, such as a tubular shape with finite thickness for zinc-blend nanowires [32] and a three-dimensional cylinder shape for zinc-blend [33] and for wurtzite nanowires [34]. For a tubular shape, two-dimensional electron gas is rolled up and a band bending occurs due to the surface-states-pinning, while for three-dimensional cases they do not include the effect of surface-states-pinning nor band-bending in the vicinity of the surface, thus assuming that electron transport is three-dimensionally diffusive.

In contrast, apart from estimation by electrical transport theory and experiments, the spin-orbit interaction can be theoretically obtained by multi-band **k·p** model. Moreover, for III-V semiconductors, it is known that a simpler equation based on the conduction band approximation can effectively deduce the spin-orbit coupling using a so-called Kane parameter [35]. However, since the Kane parameter, which includes couplings between conduction band and valence bands, does not consider coupling among valence bands, the spin-orbit coupling obtained from the effective equation happens to be inconsistent with that obtained from multi-band **k·p** model for specific crystal structures.

Recently, Escribano et al. modified the Kane parameter for the results to be consistent with 8 band **k·p** model [31]. They demonstrated that, by assuming quasi-one-dimensional electron distribution, the same form of the effective equation but using the modified parameter is able to reproduce the results of the various previous measurements [6–13] that were analyzed by one-dimensional diffusive model reported in Ref. [9]. The reason for good match was explained by Ref. [31], which states that the diffusive model in Ref. [9] includes all the contribution from all the sub-bands in a nanowire. This indicates that, even though the models are differently structured, the estimated spin-orbit coupling agrees well between the microscopic band calculation and the weak antilocalization model, when the electron propagation is assumed in the same way.

Based on this background, we reconsider a model that is available to analyze our experiments. Our wurtzite InAs nanowire with mobility approximately ranging from 100 - 1400 $cm^2$/Vs (i.e. diffusive transport) includes interface states and surface states, both of which generally arise at InAs semiconductor surface. Such surface states pin the Fermi level inside the semiconductor, then making the band bending in the vicinity of the surface [11,32,36]. Therefore we currently consider that, for our wurtzite InAs nanowire, electron might propagate in a tubular shape, as is modeled for zinc-blend InAs nanowire in Ref. [32]. However, a weak antilocalization model with a tubular-shape electron distribution has not been reported for wurtzite InAs nanowire. Therefore we plot the Rashba coupling constant extracted from our data using the one-dimensional diffusive model in Ref. [9], as is shown in

Fig.4(b).

We also plot calculation reported in Ref. [31], which nicely reproduces the results of our previous GAA-MOSFET. In addition, when this plot is horizontally offset, we find an approximate agreement about the gradient of $\alpha_R$ vs $V_g$ between the calculation and the experimental data for the ZnO-GAA MOSFET.

This can be reasonably explained by a simple estimation. When we consider a cylinder capacitance model, the peak electric field $E_p$ at InAs nanowire surface is given by $E_p \propto (C/L_g) \times (1/W) \times \Delta V_g$ ($C$: capacitance of the device, $\Delta V_g$: increment in $V_g$ from the flat-band condition). The term of $(C/L_g) \times (1/W)$ ($\propto E_p/\Delta V_g$) only differs by 16 % between our previous metal-GAA device and ZnO-GAA device. Thus, it seems reasonable that the $V_g$ dependence of $\alpha_R$, being proportional to $E_p/\Delta V_g$ in a rough estimation, is approximately the same between the devices.

In conclusion, we demonstrate that a GAA -type InAs nanowire using *in-situ* grown gate stack of $Al_2O_3$/ZnO largely tune the spin-orbit interaction with a small gate bias. This will be useful to future nanoscale spin-orbitoronics and optoelectronics.

Acknowledgement

This work was supported by JSPS KAKENHI Grant Number JP20H02562.

The data that support the findings of this study are available from the corresponding author upon reasonable request.

Figures

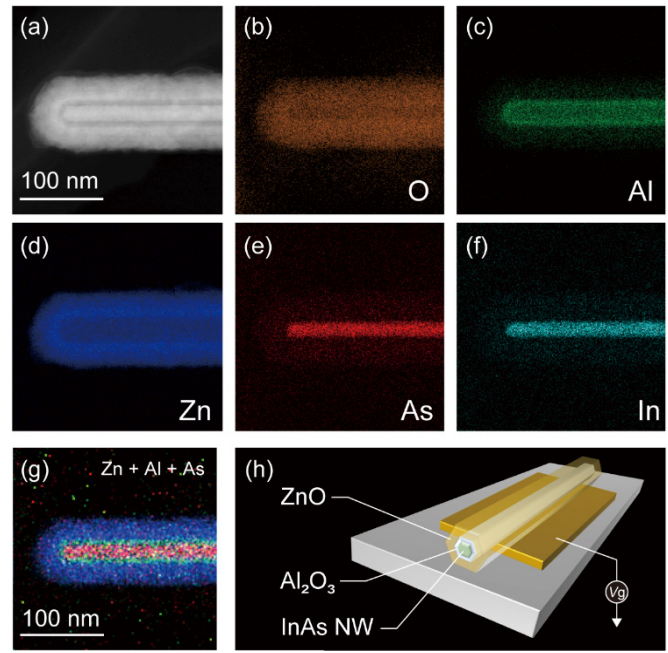

Figure 1 (a)-(g) Scanning TEM images. (a) High-angle annular dark field image. The scale bar is 100 nm. (b) to (f) EDS images taken for each element. Every image shows no notable unintended migration of elements. (g) Overlapping of the images (c), (d) and (e). The scale bar is 100 nm. (h) The schematic illustration of our gate-all-around nanowire FET.

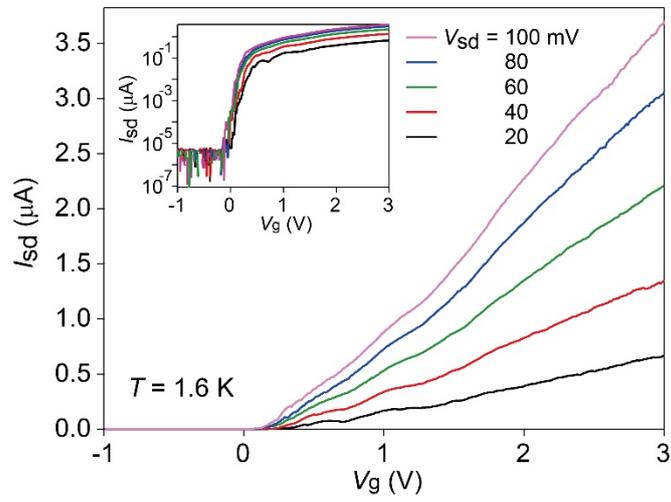

Figure 2 Source-drain current as a function of gate voltage, which is measured for various source-drain voltages. The inset shows a semi-log plot.

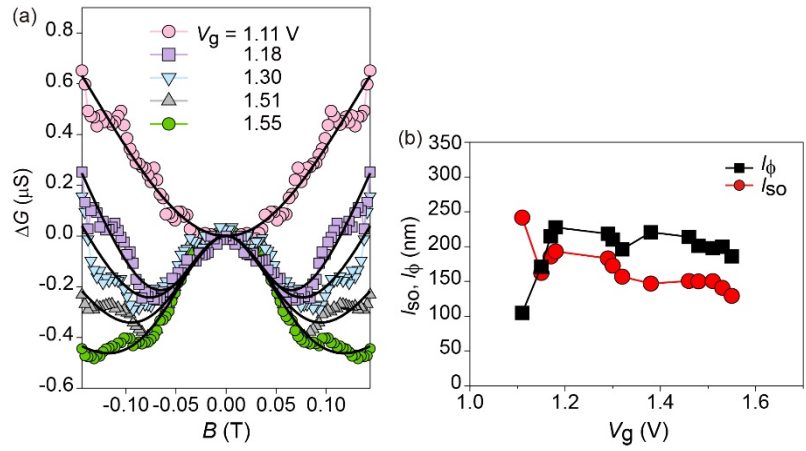

Figure 3  (a) magnetoconductance as a function of magnetic fields. The lines are the fitting curves obtained with a one-dimensional diffusive model. (b) The spin-orbit length and phase relaxation length vs gate voltage.

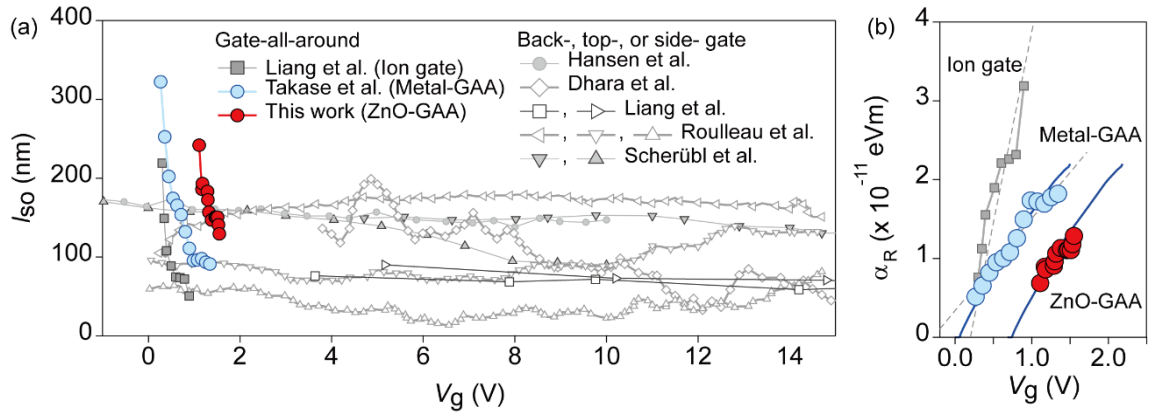

(double-column figure)

Figure 4 (a) Comparison of $l_{so}$ among various types of InAs nanowires [6,7,9–13], which are plotted as a function of gate voltage. (b) The Rashba coupling constant vs. gate voltage. The dashed lines are the guides to the eye. The blue line along the metal-GAA data is extracted from Ref. [31]. The other blue line is a horizontally offset one to compare the ZnO-GAA data.